\documentclass{ws-procs9x6}

\def\jipsi{\ensuremath\mathrm{J}/\psi}
\newcommand{\D}{\ensuremath\mathrm{D}}
\newcommand{\Db}{\ensuremath\overline{\mathrm{D}}{}}
\newcommand{\B}{\ensuremath{\mathrm{B}}}
\newcommand{\K}{\ensuremath{\mathrm{K}}}

\newcommand{\Xmass}{\ensuremath{\mathrm{X(3872)}}}
\newcommand{\Xpmass}{\ensuremath{\mathrm{X(3941)}}}
\newcommand{\e}{\ensuremath{\mathrm{e}}}

\def\tso{\ensuremath{{}^3\mathrm{S}_1}}
\def\tdo{\ensuremath{{}^3\mathrm{D}_1}}

\def\spo{\ensuremath{{}^1\mathrm{P}_1}}
\def\tpz{\ensuremath{{}^3\mathrm{P}_0}}
\def\tpo{\ensuremath{{}^3\mathrm{P}_1}}
\def\tpt{\ensuremath{{}^3\mathrm{P}_2}}

\begin{document}

\title{Speculations in hadron spectroscopy\footnote{\uppercase{P}reprint  \uppercase{LPSC}-04-110, hep-ph/0412252}\footnote{%
\uppercase{I}nvited talk at the \uppercase{SPIN} 2004 \uppercase{C}onference, \uppercase{T}rieste, \uppercase{O}ctober 10--16, 2004}}

\author{Jean-Marc Richard}

\address{Laboratoire de Physique Subatomique et Cosmologie,\\ 
        Universit\'e Joseph Fourier-- CNRS-IN2P3,\\
        53, avenue des Martyrs, 38026 Grenoble cedex, France\\ 
E-mail: jean-marc.richard@lpsc.in2p3.fr}

\maketitle

\abstracts{
A selected survey is presented of the recent progress in hadron spectroscopy. This includes spin-singlet charmonium states, excitations of charmonium and open-charm mesons, double-charm baryons, and pentaquark candidates. Models proposing exotic bound states or resonances are reviewed. The sector of exotic mesons with two heavy quarks appears as particularly promising.}

\section{Introduction}
A dramatic revival of hadron spectroscopy has been observed for several months. In high-energy experiments, a larger part of the analysis effort is devoted to the search for new hadrons. Theorists also rediscover how challeging are the issues raised by confinement in QCD.

Some problems have been here for years. Progress is sometimes slow but effective: for instance, the data accumulated by the annihilation experiments at LEAR shed valuable light on the sector of scalar mesons, where an excess of states is observed, as compared with a counting solely based on orbital excitations of  $({\bar q}q)$ or $({\bar s} s)$ configurations. Here $q$ denotes any light quark $u$ or $d$.
The progress is more rapid for the singlet-states of charmonium, with the recent discovery of the long-awaited $\eta'_c$ and $h_c$ states.
Another example, now on the theory side, deals with the Roper resonance in the spectrum of nucleon excitations. In conventional approaches, it comes too high, above the negative-parity  excitations. Simple models with Goldstone-boson exchange, or more elaborate lattice simulations\,\cite{Dong:2003zf} with low value for the mass of the light quarks, clearly show that the level ordering is due to chiral dynamics.

Other problems were bothering only a group of pioneers, and have been unveiled by the recent experimental findings. Immediately comes to mind the existence of a low-lying pentaquark antidecuplet, predicted by chiral-soliton models. Another issue deals with stable or narrow $\D\Db{}^*$  molecules.
Also hot is the question of parity partners lying close above some ground-state configurations, and remarkably illustrated by the so-called $\D_{s,J}$ resonances. 

These  sectors will now be discussed, and then I shall comment on the possibility of predicting stable, exotic multiquark states from our present knowledge of quark dynamics. I apologise in advance to be unable, due to the lack of space,  to cite all the relevant references to new results and new ideas which are proposed almost every day.
%%%%%%%%%%%%%%%%
\section{Charmonium}
%%%%%%%%%%%%%%%%
Thanks to an energetic coordination, the 
``Yellow Report'' of the \emph{Quarkonium Working Group}\,\cite{qwg} is now finished, and provides a summary of experimental and theoretical aspects, with many references. For this spin conference, it seems appropriate to focus on the spin-dependent forces beween quarks.  The non-observation in $\e^+\e^-$ of any \tdo\ of $(b\bar{b})$ confirms that if the $\psi''$ state of  the charmonium family is seen formed in $\e^+\e^-$, this is due to its $\tso$ mixing induced by tensor forces, suppressed by 
$(m_c/m_b)^2$ for Upsilon, and to the  particular coupling induced by the neighbouring $\D\Db$, a situation without analogue for the 2S level of $(b\bar{b})$. The  $\tso\leftrightarrow\tdo$ orbital mixing also influences the decay properties of $\psi'$ and $\psi''$ states.

While radial ($\psi'$, $\psi'''$, \dots) or orbital ($\chi_J$, $\psi''$, \dots) excitations of the $\jipsi$ have been identified rather early in $\e^+\e^-$ experiments and confirmed in production and $\bar{\rm p}{\rm p}$ formation experiments, the spin-singlet partners have been more elusive. This was expected, but not to that level.

The $\eta_c\,(1S)$ was first found 300 MeV below the $\jipsi$, before a more reasonable hyperfine splitting of about 120\, MeV was measured. Its radial excitation, $\eta'_c$, or $\eta_c\,(2S)$ was searched for in several experiments, including $\psi'$ radiative decay, $\gamma\gamma$ fusion or $\bar{\rm p}{\rm p}$ formation. The embarrassing absence of reliable $\eta'_c$ is over now, with the observation by the Belle collaboration of $\eta'_c$ in two different measurements\,\cite{qwg}. The first one is the double-charm production in $\e^+\e^-$ collisions, that is to say, $\jipsi + X$. The Zweig rule strongly suggests that $c\bar{c}$ recoils against $c\bar{c}$. There is, however, some debate on the cross-section for these double-charm events.  The second Belle analysis deals with \B\ decay. The final state $\K\K\K\pi$ exhibits a peak in the $\K\K\pi$ mass spectrum. The $\eta'_c$ has also been seen in other experiments\,\cite{qwg}. The $\eta'_c$ is closer to $\psi'$ than expected in simple charmonium models. This is probably due to the  coupling to the open-charm threshold\,\cite{qwg}. The $\eta_c'$ does not link to $\D\Db$, due to its pseudoscalar nature, but the $\psi'$ does, and is pushed down. 

The $\spo$ state of charmonium, $h_c$,  was searched for  actively.  It was suspected in the R704 experiment at CERN ISR, and seen for a while in the Fermilab $\rm{\bar p}p$ experiment, not confirmed in the first analysis of further runs, and  eventually seen by looking at another final state, and also detected at CLEO\,\cite{qwg,Tomaradze:2004sk}. The mass lies, as  expected, at the centre of gravity of \tpz, \tpo\ and \tpt\ triplet states. 
This means that, in a  perturbative analysis of spin forces within the potential models, there is no spin--spin interaction acting on the P-state multiplet.
There are presumably several corrections to this simple analysis, which tend to cancel out on the whole.
Higher-order QCD corrections give some corrections to the $\spo$ mass, and, indeed, the spin--spin potential obtained from lattice calculation is not strictly of zero range.  However, if the spin--orbit and tensor forces are treated non perturbatively, the $\spo$ should not be compared to the naive centroid  $(\tpz+3\tpo+5\tpt)/9$, but to an improved average which is higher by a few MeV.  The energy of triplet states is, indeed, not a linear, but a concave function of the coefficients of the spin--orbit and tensor components.

Higher states of charmonium might reveal new structures. The $\psi(4040)$ has an interesting history in this respect. It was found to have preferential decay into $\D\Db^*$ (an implicit ${}+ {\rm c.c.}$ is implied here and in similar circumstances). This suggested a molecular description of this state\,\cite{Voloshin:1976ap}. In fact, the groups at Orsay  and  Cornell  understood that the unorthodox pattern of  the branching ratios to $\D\Db$, $\D\Db^*$ and $\D^*\Db^*$ is due to the radial structure of $\psi(4040)$, as a mere $(c\bar{c})$ state\,\cite{LeYaouanc:1977gm}. The spatial wave function has nodes. In momentum space, there are also oscillations. Hence, if a decay calls for a momentum whose probability is low, it is suppressed.

A more recent state in the charmonium mass-range is the $\Xmass$\,\cite{Aubert:2004ns}, seen in several experiments, as a clear signal above  the background, and hence  considered as very safely established. The $\Xmass$ is not seen in two-photon production at CLEO, thus constraining the possible quantum numbers. Though experts have learned to be careful from the lesson of the $\psi(4040)$, a molecular interpretation of this state is very tempting, as 
it lies almost exactly at the $\D^0\Db^{*0}$ threshold, and 
none of the charmonium assignment (D-state, radially excited P-state, etc.) survives scrutiny, although the production is rather reminiscent of the pattern observed for usual charmonium states. 
Such a molecule was predicted  on the safe ground of the Yukawa interaction between two hadrons, and recently generalised to bound states of charmed baryons\,\cite{Tornqvist:2003na}.

More recently, another state has been seen in this region, let us call it $\Xpmass$, decaying into $\jipsi+\omega$. As noted in the paper revealing its existence\,\cite{Abe:2004zs}, it bears some properties of the charmonium hybrid, as predicted by  Giles and Tye, Mandula and Horn, Hasenfratz et al., and many others\,\cite{Giles:1977mp}.  Hybrid mesons have an explicit constituent gluon, and this allows for exotic quantum numbers that are not permitted for $(Q\overline{Q})$, and also leads to  supernumerary states with ordinary quantum numbers. Though the first experimental candidates for hybrid mesons were proposed in the light sector, the heavy sector offers the advantage of  a cleaner knowledge of ordinary  $(Q\overline{Q})$, on the top of which the unusual states are better singled out.
\section{Open-charm mesons}
States with $(c\bar{s})$ flavour content were found at the Babar experiment of SLAC, and confirmed at CLEO, Fermilab, etc.~\cite{Bondioli:2004te}.
The masses, $2317$ and $2458\;\mathrm{MeV}$, can be compared with 1968 for the pseudoscalar and $2112\;\mathrm{MeV}$ for the vector ground state of $(c\bar{s})$. This is rather low for the lowest P-state, from our present understanding of spin--orbit forces. 
The small widths are also rather intriguing.
Views  about these states schematically fall into two categories: \\[-15pt]
\begin{enumerate}
\item
These new states are  understood as the chiral partners of the ground-state multiplet.  Some authors insist on that this is not an ad-hoc explanation just for these two states, but a recurrent phenomenon for light quarks surrounding a heavy core \cite{Bardeen:2003kt}.
\item
Barnes et al.\ \cite{Barnes:2003dj} and several other authors proposed a four-quark interpretation,  $(cq\bar{s}\bar{q})$,  with, however, a possible breaking of isospin symmetry.
\end{enumerate}\vskip -4pt
Further investigation is needed to decide whether the new $\D_{s,J}$ states are just the usual excitations, shifted in mass and made narrower than expected, or supernumerary states.

This summer,  a new state, $D_{s}(2632)$, was announced by the Selex collaboration, again rather narrow, and decaying more often into $\D_s\eta$ than into $\D\mathrm{K}$~\cite{Evdokimov:2004iy}. It has not been confirmed \cite{Aubert:2004ku}. Several interpretations have been proposed for the  $D_{s}(2632)$, among them a four-quark structure $(cq\bar{s}\bar{q})$, somewhat reminiscent\,\cite{Nicolescu:2004in} of the baryonium states proposed in the late 70's.
%
%%%%%%%%%%%%%%
\section{Double-charm baryons}
%%%%%%%%%%%%%%%
According to the results shown by Selex at the INPC conference,  the lowest $\Xi^+_{cc}(ccd)$ state is now seen in two different weak-decay modes. Problems  remain: Selex is the only experiment having seen these baryons, yet; the isospin splitting between this $\Xi^+_{cc}$ and the lowest $\Xi^{++}_{cc}$ candidate is larger than expected; the puzzling excitations about 60 MeV  above the ground state need confirmation with higher statistics.

Anyhow, the  Selex results have stimulated further studies on hadrons with two heavy quarks. The $(QQq)$ baryons are perhaps the most interesting of ordinary hadrons, as they combine in a single object two extreme regimes: the slow motion of two heavy quarks in an effective potential generated by light degrees of freedom, as in charmonium or Upsilon systems; the ultra-relativistic motion of a light quark around a heavy colour source, as in $\D$ and $\B$ mesons.
The large $Q/q$ mass ratio, which implies a hierarchy $r(QQ)\ll r(Qq)$ of mean radii, suggests  valuable approximation schemes\,\cite{qwg} such as a diquark--quark picture or a Born--Oppenheimer treatment of $(QQq)$.

As pointed out\,\cite{Bardeen:2003kt}, one expects chiral partners of $(QQq)$ baryons, similar  to the chiral partners of $(Q\bar{q})$ mesons. However, a universal spacing of about $300\;$MeV is empirically observed between any hadron and its chiral partner. Hence the small spacing suggested by the Selex results is difficult to understand.
%%%%%%%%%%%%%%%%
\section{Pentaquarks}
%%%%%%%%%%%%%%%%%
A baryon with charge $Q=+1$ and strangeness $S=+1$, i.e., minimal quark content $(\bar{s}uudd)$ has been seen at the Spring8 facility in Japan, and in several other experiments.  However, this state is not seen in a number of high-statistics experiments with a very good particle identification, and for some of them, with very good record in hadron spectroscopy. Hence, the status of the $\theta^+$ is more than shaky\,\cite{Rossi:2004rb}. 

The situation is even worse for the $\Xi^{--}$ member of the putative antidecuplet, since it has been claimed only by a fraction\,\cite{Alt:2003vb} of the NA49 collaboration, and not seen in any other analysis\,\cite{Adamovich:2004dr} looking for it.
It has also been proposed that some baryon resonances with ordinary quantum numbers contain a large fraction of pentaquark configurations, to explain their intriguing properties. This concerns, e.g., the Roper resonances and the $\Lambda(1405)$.

On  the theory side, the situation is also rather confusing, even for experts. The pioneers on chiral soliton dynamics, and in particular, the Skyrmion model, made the remarkable prediction\,\cite{Diakonov:2004ie} of an antidecuplet ($\overline{10}$) of baryons on almost the same footing as the familiar octet ($\mathrm{N}$, $\Lambda$, $\Sigma$, $\Xi$) and decuplet ($\Delta$, \dots, $\Omega^-$). The positive-parity assignment is crucial in this picture, and is confirmed in simpler constituent models\,\cite{Stancu:2003if} mimicking the chiral effects by spin--flavour terms.

Moreover, lattice QCD calculations and QCD sum rules have seen either negative parity, or positive parity, or both, or no state at all. Critical surveys\,\cite{Sasaki:2004vz} exist. As in ordinary few-body quantum mechanics, it is essential to demonstrate a clear separation between the continuum dissociation threshold and possible resonances on the top of this background.

Constituent models have been worked at very hard, perhaps too hard, to accommodate the announced pentaquark. In particular, astute diquark or even triquark clustering has been deviced, that, once accepted, naturally leads to pentaquark\,\cite{Jaffe:2004ph}. Note that early promoters of the diquark in soft physics do not fully endorse this use of diquark\,\cite{Lichtenberg:2004tb}.

In the past, diquarks were successfully advocated to explain why mesons and baryons exhibit the same Regge slope, but the $[qq]-q$ clustering of orbitally excited baryons was later demonstrated\,cite{Martin:1985hw}  to occur in a large class of potential models. However, $[qq]-[\bar{q}\bar{q}]$ clustering was also postulated for four-quark states, leading to the prediction of many baryonium states, but these states were never confirmed, and,  to my knowledge, no serious four-body calculation was ever attempted to support this type of clustering.

In most of the models explaining the light pentaquark with strangeness $S=+1$, a heavy $(\overline{Q}qqqq)$ version, tentatively more stable against dissociation,  exists. Heavy pentaquarks have already a long history. A first candidate was proposed in 1987 independently in a paper by the Grenoble group, where the word ``pentaquark'' was seemingly used for the first time, and another one by Lipkin \cite{Gignoux:1987cn}. It consists of $(\overline{Q}qqqq)$, with the light--strange sector $q^4$ forming a flavour triplet of SU(3). The parity is negative in the original model, very much inspired by Jaffe's H$(uuddss)$, the binding being due to attractive coherences in the chromomagnetic interaction.  
%%%%%%%%%%%
\section{Stability of multiquark states}
%%%%%%%%%%%
It is sometimes claimed, even in otherwise  remarkable papers\,\cite{Carlson:1991zt}, that the potential models produce a large number of stable or metastable states, and hence, are ruled out by the scare evidence for such states.
However, the reverse is true: potential models, supplemented by the most current ansatz for the colour dependence of the interaction, predict that multiquarks do \emph{not} bind below their dissociation threshold, except for the rare configurations where a special coherence  benefits  the collective quark system, and not  its decay products.
Stability of few-body systems should, indeed, be seen as a competition between a collective behaviour and  preference for dissociation.  

For a Hamiltonian with a pairwise potential $V\propto\sum\lambda^{\rm c}_i.\lambda^{\rm c}_j \, v(r_{ij})$, with the usual (though questionable) colour dependence, it is found that baryons are heavier \textsl{per quark} than mesons\,\cite{Nussinov:1999sx}, and further, for equal mass quarks, that $(qq\bar{q}\bar{q})\ge 2(q\bar{q})$. Some effects, however, might tend to reverse this inequality, and its analogues such as $(qqqq\bar{q})\ge (qqq)+(q\bar{q})$, $(q^6)\ge 2(q^3)$, etc. This includes, for instance, spin--colour forces, or spin--flavour forces.

Another, simpler, effect deals with mass differences. In a (flavour independent) confining potential, heavy quarks take better advantage of the binding, exactly in the same way as a particle  in an harmonic well  has  its  binding energy  $\propto\mu^{-1/2}$ decreasing when its mass $\mu$ increases. This makes $(QQ\bar{q}\bar{q})$ taking advantage of the heavy--heavy interaction, that is absent in the threshold made of two separate $(Q\bar{q})$ mesons.

There is also something special with light quarks that cannot be extrapolated from potential models tuned to heavy quarks. This goes beyond the necessary account for relativistic effect. The following exercise is instructive: take a simple linear potential $\lambda r$, or an improved $\lambda r - a/r$; use relativistic kinematics, and tune the parameters ($\lambda$, $a$, quark masses), to reproduce the spin-average levels of charmonium and $\D$ meson. Then solve for $(qqq)$, $(qqQ)$, $(qQQ)$, $(QQ\bar{q}\bar{q})$, etc., so that the heavy--heavy effect is automatically taken into account. It is found  that $(ccq)$ comes out compatible with the Selex mass (if room is left for hyperfine effects), and that $(QQ\bar{q}\bar{q})$  becomes lower than $2(Q\bar{q})$ if quark-mass ratio $M/m$ is large enough. 

This model, however, will miss some of the light--light attraction, by overestimating $(Qqq)$ by about 150 MeV, and $(qqq)$ by 500 MeV. An additional  dynamical ingredient, which   breaks flavour independence, should be implemented. It is responsible for the nuclear forces among hadrons containing light quarks, and is approximately realised in empirical models with Yukawa-type scalar--isoscalar exchange between light quarks. Altogether, if one computes the $(cc\bar{q}\bar{q})$ mass with a model compatible with the $(c\bar{c})$ and $(c\bar{q})$ spectra, and adds about 100 to 150 MeV extra attraction for the light-quark pair,  one finds a bound state that decays weakly. 

In short, the $(cc\bar{q}\bar{q})$ configuration seems one of the most promising candidates for a stable multiquark state, since it benefits both from the heavy--heavy and light--light effects.
%%%%%%%%%%%%%%%%%%%
\section{Conclusions}
%%%%%%%%%%%%%%%%%%%
The last months have been very exciting for hadron spectroscopy. The $\D_{s,J}$ resonances, the heavy baryons, the $\Xmass$, and the pentaquark candidates stimulated interesting studies on confinement dynamics.

The discovery of the $\eta'_c$ shows that new means of investigation can solve the old problems, and it is hoped that the missing baryons and quarkonia will also be found with an appropriate production mechanisms.  

For many years, hadrons with multiquark structure, or constituent glue, or revealing the power of chiral symmetry, have been searched for. Now, they are emerging perhaps too suddenly, and we would be rather embarrassed if all the recent candidates survive careful experimental scrutiny. We have to wait for the current next wave of experiments and analyses, especially concerning the controversial pentaquark.

Meanwhile, theorists should also refine and improve their tools. The history of poly-electrons  is in this respect rather instructive. Around the year 1945, Wheeler proposed several new states, in particular the $(\e^+,\e^+,\e^-,\e^-)$, as being stable if internal annihilation is neglected.  In 1946, Ore published an article where he concluded that stability is very unlikely, on the basis of a seemingly-solid variational calculation borrowed from a nuclear-physics picture of the $\alpha$-particle as a four-nucleon system. 
Hylleraas, however, suspected that the trial wave function was not suited for long-range forces. Today, most of us, in similar circumstances, would rush to their computer and post a criticism on the web. These gentlemen, instead, combined their efforts, and in 1947, published a very elegant and rigorous proof of the stability \cite{Whe46}.
%%%%%%%%%%%%%%%%%
\subsection*{Acknowledgments}
I would  like to thanks Franco Bradamante and his colleagues for this beautiful conference. I benefitted from several discussions with Ica Stancu,   useful remarks on the manuscript by Muhammad Asghar, and informative correspondence with  Harry Lipkin and Stephan Paul.
%%%%%%%%%%%%%%%%

%\vskip .5cm
\newpage
%%%%%%%%%%%%%%%%%%%%%%%%%%%%%%%%%
%%%%%%%%%%%%%%%%%%%%%%%%%%%%%%%%%%
\textbf{Question by O.\ Teryaev}
In the QCD sum-rule approach of Ioffe and Oganesian, the small width of the $\theta^+$ is explained by chiral invariance. If so, one may expect a large non-forward transversity distribution for the transition from nucleon to $\theta^+$, which may lead, in turn, to the electroproduction of $\theta^+$ accompanied by transversely polarized $\mathrm{K}^*$ and longitudinally polarized $\rho$, in analogy to $\rho_{\rm L}\rho_{\rm T}$ production studied by Ivanov, Pire, Szymanowski and myself, which cannot be explained by the small width.

\textbf{Answer by J.-M. Richard}
Yes, this is a very interesting remark by Ioffe and Oganesian, that certain transitions are suppressed by chiral symmetry considerations. Similar considerations have been proposed by Melikhov and Stech, and others\,\footnote{See, e.g., Matheus and Narison\,\protect\cite{Sasaki:2004vz}}. I also agree with your second point. More generally, spin observables are very useful to single out resonances whose signal is elusive in integrated cross sections.

\vskip .5cm
\textbf{Question by Ken Imai}
It is important to confirm $\theta^+$ with high statistics and high precision. Experiments are being done at Spring8 and Jlab. The preliminary data from Spring8 have confirmed the previous result with higher statistics. We are now preparing an experiment to study $\theta^+$ with a $\mathrm{K}^+$ beam with 1\,MeV resolution at KEK-PS. Since this is a spin symposium, I would like to comment that a measurement of polarization of the final proton, or an experiment with a polarized beam is important for determining the spin and parity of $\theta^+$.

\textbf{Answer by J.-M. Richard}
I agree with you both on the need of better statistics and the importance of spin observables for extracting the quantum numbers. When one reads the \textsl {Review of Particle Properties}, one is astonished to discover that for many hadrons, the quantum numbers are not determined by unambiguous measurements but simply deduced from the most plausible quark-model assignment. For the exotics whose structure is debated, the spin and parity have to be measured.

\vskip .5cm
\textbf{Question by Hikaru Sato, KEK, Accelerator Division}
Are there some interesting physics issues, in particular for hadronic interaction and hadron spectroscopy, accessible with a 50\,GeV beam of polarized protons?

\textbf{Answer by J.-M. Richard}
Yes, the potential of such machines is extremely rich, thanks  in particular to the variety of secondary beams they can produce. In the particular sector of spectroscopy, these machines will however face a   competition from beauty factories.

\end{document}